# A DESIGN OF AN AUTONOMOUS MOLECULE LOADING/TRANSPORTING/UNLOADING SYSTEM USING DNA HYBRIDIZATION AND BIOMOLECULAR LINEAR MOTORS

*Satoshi Hiyama*[†], *Yasushi Isogawa*[‡], *Tatsuya Suda**[†], *Yuki Moritani*[†], *and Kazuo Sutoh*[‡]

[†] Network Laboratories, NTT DoCoMo, Inc.
[‡] Department of Life Sciences, the University of Tokyo
* Information and Computer Science, University of California, Irvine

## ABSTRACT

This paper describes a design of a molecular propagation system in molecular communication. Molecular communication is a new communication paradigm where biological and artificially-created nanomachines communicate over a short distance using molecules. A molecular propagation system in molecular communication directionally transports molecules from a sender to a receiver. In the design described in this paper, protein filaments glide over immobilized motor proteins along preconfigured microlithographic tracks, and the gliding protein filaments carry and transport molecules from a sender to a receiver. In the design, DNA hybridization is used to load and unload the molecules onto and from the carriers at a sender and a receiver. In the design, loading/transporting/unloading processes are autonomous and require no external control.

## 1. INTRODUCTION

Molecular Communication [1]-[2] is a new and interdisciplinary research area that spans the nanotechnology, biotechnology, and communication technology. Molecular communication allows biological and artificially-created nanomachines to communicate over a short distance using molecules. In molecular communication, a sender nanomachine encodes information onto molecules (called information molecules). Information molecules are then loaded onto carrier molecules and propagate to a receiver. The receiver, upon receiving the information molecules, reacts biochemically to the incoming information molecules.

Molecular communication is inspired by the observation that in the biological systems, communication is typically done through molecules [3]-[5]. Using molecules in communication has not yet been, however, applied to controlled communication between nanomachines. The current research effort in nanotechnology and biotechnology focuses on observing and understanding existing biological systems (e.g., observing and understanding how communication is done within a cell or between cells). Molecular communication extends the current effort to include design of nanomachine communication systems based on biological communication mechanisms.

Molecular communication is a new communication paradigm, and as such, it requires research into a number of key areas. Key research challenges include, among others, 1) design of a sender that generates information molecules, encodes information onto the generated information molecules, and emits the information molecules in a controlled manner, 2) design of a molecular propagation system that directionally transports information molecules from a sender to a receiver in an artificially created aqueous environment without external control, and 3) design of a receiver that selectively receives information molecules, decodes the information encoded in the received information molecules, and biochemically reacts to the received information.

This paper focuses on one of the key research issues, namely, the design of a molecular propagation system that directionally transports the information molecules from a sender to a receiver in an artificially created aqueous environment without external control. Note that this directional transport takes place in the presence of a noise (i.e., Brownian motion) in the transport environment, and thus, the design presented in this paper uses biomolecular linear motor systems [3]-[6] to directionally transport information molecules.

Biomolecular linear motor systems consist of motor proteins (i.e., kinesins, dyneins, and myosins) and their protein filaments (i.e., microtubules and actin filaments). In Biomolecular linear motor systems, motor proteins may move along immobilized protein filaments, or protein filaments may glide over immobilized motor proteins. Biomolecular linear motor systems, in which motor proteins move on immobilized protein filaments, can directionally transport cargoes such as vesicles and organelles along the protein filaments *in vivo* (i.e., within eukaryotic cells) [3]-[6] and also *in vitro* (i.e., in an artificially created aqueous environment) [7] in the presence of noise. However, it may be difficult with this approach to align polarity (i.e., plus- and minus-ends) of protein filaments into preconfigured network topologies of protein filaments in an artificially created aqueous environment. On the contrary, biomolecular linear motor systems, in which protein filaments glide over immobilized motor proteins, has been extensively studied in an artificially created aqueous environment [8]-[13] and may present a variable solution to the design of a molecular propagation system to directionally transport the information molecules from a sender to a receiver. Although the gliding direction of protein filaments is naturally random, recent work has demonstrated that the gliding direction of protein filaments on a preconfigured network topology of microlithographic tracks can be successfully controlled [8]-[9],[13]. Such gliding protein filaments with controlled movement can act as carrier molecules to transport information molecules from a sender to a receiver. In the design of a molecular propagation system described in this paper, thus, protein filaments (carrier molecules) glide over immobilized motor proteins along microlithographic tracks to transport information molecules.

In a molecular propagation system, it is necessary to load (at a sender) and unload (at a receiver) information molecules onto and from carrier molecules without external control. In addition, it is desirable if carrier molecules, after unloading information molecules at a receiver, are recycled and used again to load new information molecules. One approach is to use avidin-biotin binding to load information molecules onto carrier molecules [9]-[10]. This biological binding, however, may be too strong to dissociate the binding and unload information molecules at a receiver. Another approach is to use reversible inclusion phenomena caused by cyclodextrin [11] to load information molecules onto carrier molecules. This method, however, also has a downside of requiring external stimuli such as UV-light exposure to unload information molecules at a receiver, making the system non-autonomous and complex.

In the design described in this paper, DNA hybridization is used to load and unload information molecules onto and from the carrier molecules. Specifically, short single stranded DNAs (ssDNAs)-linked carrier molecules continuously glide directionally over motor proteins adsorbed to a microlithographic track surface. Information molecules at a sender are coated with long ssDNAs of the same type. Parts of the base sequences (i.e., sequences of four bases used in DNA: adenine (A), cytosine (C), guanine (G), and thymine (T)) of the long ssDNA attached to the information molecules are complementary to those of the short ssDNA attached to the carrier molecules. When short-ssDNA-linked carrier molecules pass by a sender, long-ssDNA-coated information molecules are loaded onto the gliding carrier molecules due to DNA hybridization between short ssDNAs attached to the carrier molecules and long ssDNAs attached to the information molecules. The carrier molecule-information molecule conjugates, then, directionally glide towards a receiver over immobilized motor proteins adsorbed to a microlithographic track surface. At the receiver, an array of long ssDNAs of the same type is immobilized onto the microlithographic track surface. The base sequences of the long ssDNAs at a receiver are complementary in their entirety to those of the long ssDNAs attached to the information molecules. When the carrier molecule-information molecule conjugates pass by their receiver, information molecules are unloaded from the gliding carrier molecules because partial DNA hybridization (between short ssDNAs attached to the carrier molecules and long ssDNAs attached to the information molecules) is replaced with complete DNA hybridization (between long ssDNAs immobilized onto the track surface of their receiver and long ssDNAs attached to the information molecules).

The rest of the paper is organized in the following manner. Section 2 presents a detailed design of the molecular propagation system. Section 3 presents the current status of the empirically examining the design. Section 3 also describes future research plans. Section 4 concludes the paper.

## 2. SYSTEM DESIGN

This section describes a design of an autonomous molecular propagation system that loads, transports, and unloads information molecules using DNA hybridization and biomolecular linear motors. In designing a molecular propagation system, the following design philosophy has been adopted in this paper. First, the molecular propagation system designed in this paper uses molecule-based nanometer-scale components in the biological systems. Biological components use nanometer-scale chemical energy sources such as adenosine triphosphate (ATP) with extremely low energy consumption and without requiring electrical or mechanical energy sources. As explained later in this section, the molecular propagation system uses biological components such as biomolecular linear motors and DNAs. Second, the designed molecular propagation system is a closed system, namely, a system that is autonomous without external control. As explained later in this section, the molecular propagation system achieves successive loading/ transporting/unloading of information molecules and recycling use of carrier molecules without requiring external control.

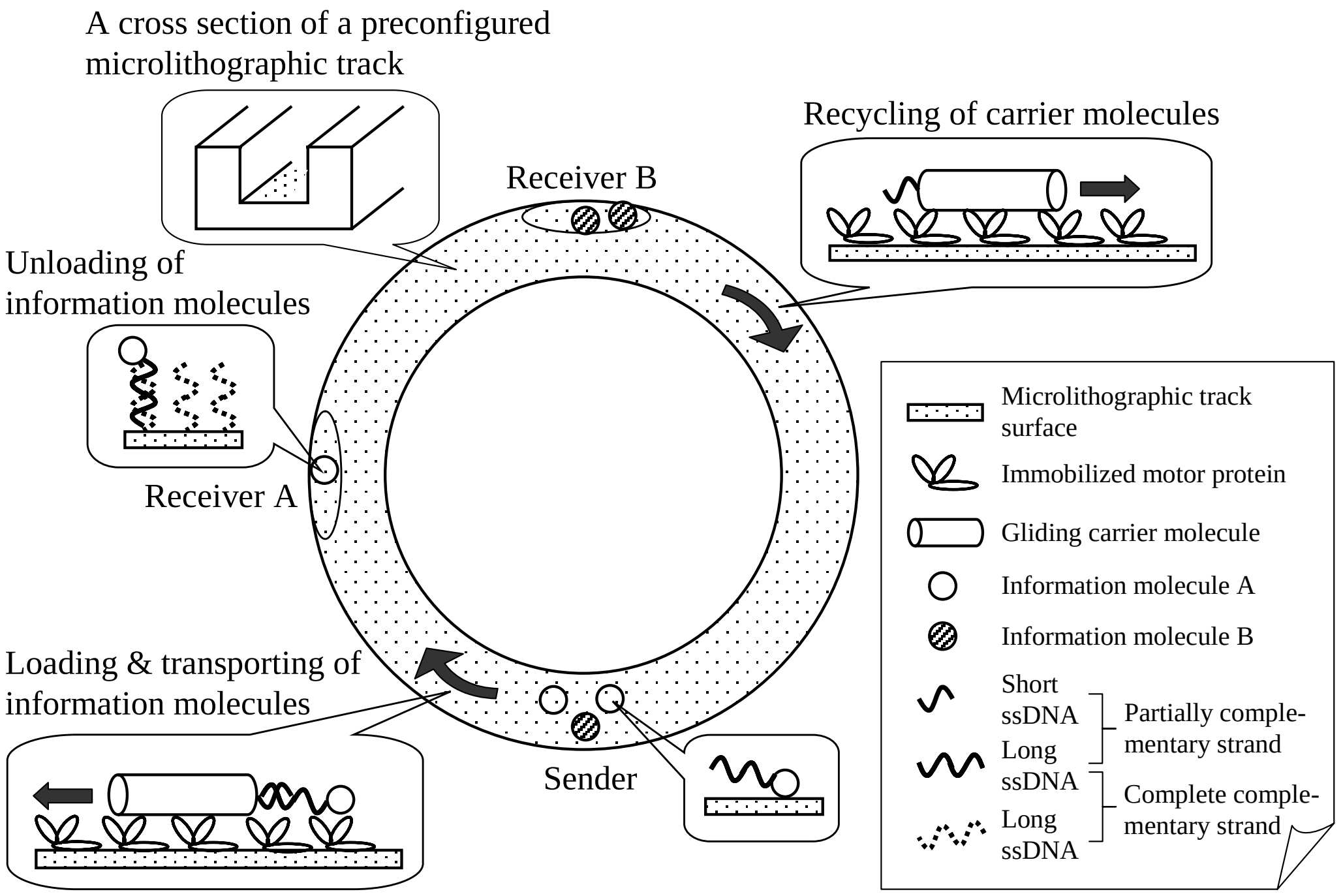

**Figure 1:** A schematic diagram of an autonomous molecular propagation system using DNA hybridization and biomolecular linear motors.

Figure 1 shows a schematic diagram of an autonomous molecular propagation system described in this paper. Key design ideas are to use protein filaments as carrier molecules of information molecules and to use DNA hybridization to load and unload the information molecules onto and from the carrier molecules. A detailed design is described below.

### 2.1. Loading of Information Molecules

In the design, short (e.g., 12 bases) single stranded DNAs (ssDNAs) of the same type are attached to a carrier molecule (i.e., a protein filament) that transports information molecules from a sender to a receiver. Note that in designing the base sequences of ssDNAs, careful attention has been paid for the structure of designed strands (i.e., not to form hairpin-loop structure) and the melting temperature ($T_m$) of designed strands (i.e., easy to hybridize at room temperature). Short-ssDNA-linked carrier molecules glide directionally over motor proteins adsorbed to a preconfigured microlithographic track surface. Information molecules are coated with long (e.g., 32 bases) ssDNAs of the same type. Parts of the base sequences of the long ssDNAs are complementary to those of the short ssDNAs attached to the carrier molecules. Long-ssDNA-coated information molecules are large enough in their sizes (e.g., micrometer-scale) such that they stay and wait to be loaded at a sender without diffusing or dispersing in the aqueous environment of a microlithographic track.

When a short-ssDNA-linked carrier molecule passes by a sender, long-ssDNA-coated information molecules are loaded onto the gliding carrier molecule due to DNA hybridization between the short ssDNAs attached to the carrier molecule and the long ssDNAs attached to the information molecules. Figure 2 shows an example of loading of an information molecule (a carrier molecule-information molecule conjugate) considered in the design described in this paper. In the example shown in this figure, parts of the base sequences (i.e., 12 bases of a short ssDNA attached to a carrier molecule and 12 bases of a long ssDNA attached to an information molecule A) bind to each other and form a double strand. Note that the rest of the base sequences of the long ssDNA attached to information molecule A (i.e., 20 bases) remains single stranded. This partial DNA hybridization is not as strong as the complete DNA hybridization and allows unloading of information molecules at a receiver. Detailed unloading mechanisms will be described later in section 2.3.

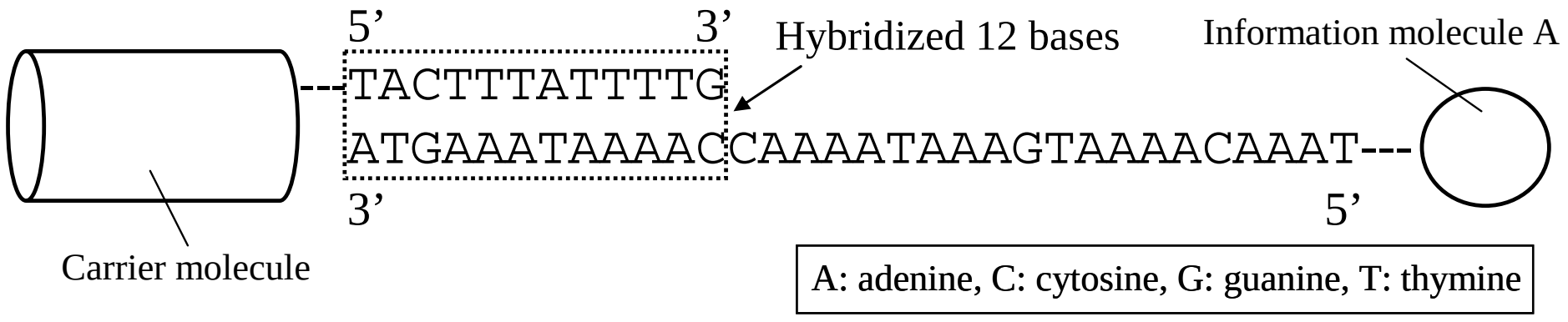

**Figure 2:** An example of loading of an information molecule.

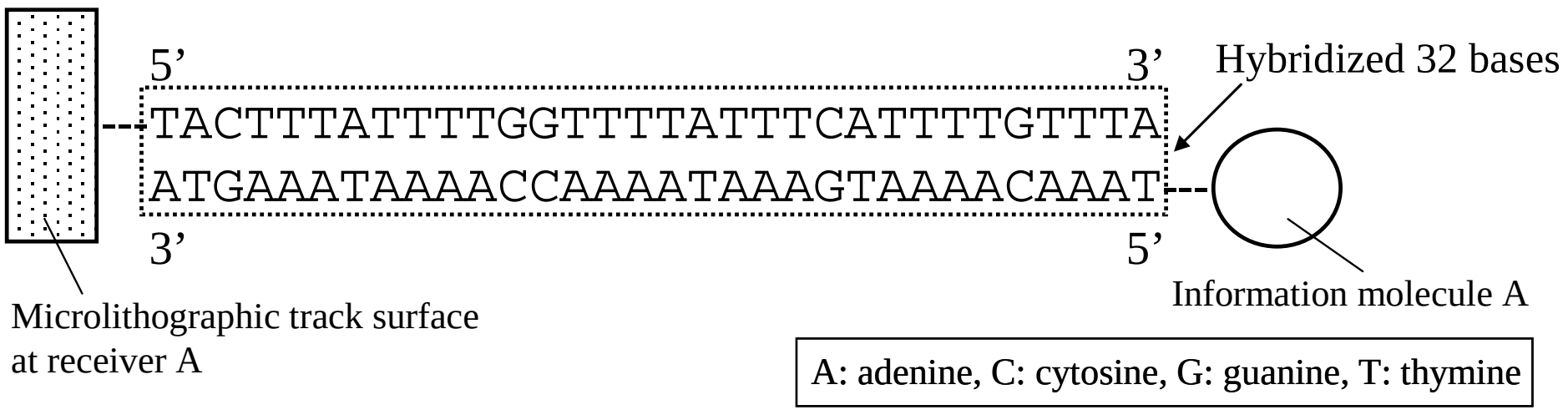

**Figure 3:** An example of unloading of an information molecule.

**2.2. Transporting Information Molecules**

The carrier molecule-information molecule conjugate directionally glides towards its receiver over motor proteins adsorbed to a microlithographic track surface. Recent work has demonstrated that the carrier molecule directionally glides along a preconfigured network topology (e.g., ring network) of a microlithographic track [8]-[9], [13], and techniques used in such existing work may be applied to control the gliding direction of a carrier molecule-information molecule conjugate.

**2.3. Unloading of Information Molecules**

At a receiver, an array of long (e.g., 32 bases) ssDNAs of the same type is immobilized onto the microlithographic track surface. The base sequences of the long ssDNAs at a receiver are complementary in their entirety to those of the long ssDNAs attached to information molecules.

When a carrier molecule-information molecule conjugate passes by its receiver, information molecules are unloaded from the gliding carrier molecule because partial DNA hybridization (between short ssDNAs attached to a carrier molecule and long ssDNAs attached to an information molecule) is replaced with complete DNA hybridization (between long ssDNAs immobilized onto the track surface of a receiver and long ssDNAs attached to an information molecule). Figure 3 shows an example of unloading of an information molecule considered in the design described in this paper. In the example shown in this figure, all the base sequences (i.e., 32 bases immobilized onto a microlithographic track surface at receiver A and 32 bases attached to an information molecule A) bind to each other and form a complete double strand. This unloading process may be initiated by the energy state transition from an unstable and high-energy state (i.e., a carrier molecule-information molecule conjugate based on the partial DNA hybridization) to a stable and low-energy state (i.e., a receiver-information molecule conjugate based on the complete DNA hybridization) that takes place naturally. Although a few of the base sequences of short ssDNAs attached to a carrier molecule still may bind to those of long ssDNAs attached to an information molecule due to the stochastic nature of DNA hybridization (i.e., chemical reactions caused by hydrogen bonds), such remaining bonds may be released by the physical force produced by motor proteins that glide the carrier molecule. Note that short-ssDNA-linked carrier molecules which unloaded information molecules continue to glide over immobilized motor proteins along the microlithographic track and may load new information molecules at a sender.

**2.4. A System with Multiple Receivers**

As described in the previous subsections, the base sequence of a long ssDNA attached to an information molecule serves as the address of an intended receiver of the information molecules. By carefully designing and using different base sequences, the design described in this paper supports a system with multiple receivers. Consider the following example. Assume that an information molecule B is coated with long (e.g., 32 bases) ssDNAs of the same type. Parts of the base sequences of the long ssDNA (i.e., 12 bases of 3’ side) are complementary to those of the short ssDNA attached to a carrier molecule (Figure 4). In the example shown in this figure, a carrier molecule-information molecule conjugate passes through receiver A without hybridizing because the single stranded base sequences of an information molecule B (i.e., 20 bases of 5’ side) are not complementary to those of receiver A at all. On the other hand, when the same carrier molecule-information molecule conjugate passes by receiver B, it unloads information molecule B because the single stranded base sequences of information molecule B (i.e., 20 bases of 5’ side) are complementary to those of receiver B, and thus, unloading process initiates.

| | |
|---|---|
| 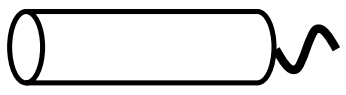 ssDNA attached on a carrier molecule | 5’ 3’ TACTTTATTTTG |
| 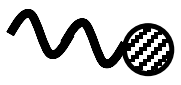 ssDNA attached on an information molecule B | 3’ 5’ ATGAAATAAAACTTTTCATTTTATTGGTTTTA |
| 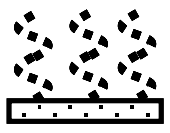 ssDNA immobilized on a track surface at receiver A | 5’ 3’ TACTTTATTTTGGTTTTATTTCATTTTGTTTA |
| 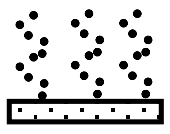 ssDNA immobilized on a track surface at receiver B | 5’ 3’ TACTTTATTTTGAAAAGTAAAATAACCAAAAT |

**Figure 4:** An example design of the base sequences for a system with multiple receivers.

## 3. CURRENT STATUS AND FUTURE RESEARCH PLANS

The authors of this paper are currently conducting biochemical experiments and empirically examining the design of a molecular propagation system described in this paper. The experiments use kinesin-driven microtubules (as carrier molecules), microbeads (as information molecules), and oligonucleotides (as short and long ssDNAs). Various binding methods of oligonucleotides and microtubules have been examined using a chemical linkage, and it is determined that succinimide ester-maleimide heterobifunctional cross-linker is most suitable for the motility of oligonucleotides-linked microtubules over the immobilized kinesins.

In the next phase of the experiments, the authors of this paper plan to examine loading and transporting of oligonucleotides-linked microbeads. In order to successfully load and transport oligonucleotides-linked microbeads, the following parameters will be varied in the planned experiments; the density of oligonucleotides, the length of the base sequence of oligonucleotides, and the gliding speed of oligonucleotides-linked microtubules.

Once the loading and transporting of information molecules are successfully examined, then unloading of oligonucleotides-linked microbeads will be examined. One difficulty at a receiver may lie in how to immobilize an array of oligonucleotides onto the kinesin adsorbed micro-lithographic track surface. In the experiments, it is required that oligonucleotides are immobilized and that they do not interfere with the functionality of kinesins. It is also required that the oligonucleotides do not to sink into the protein carpet of kinesins and caseins (commonly used proteins to prevent denaturation of kinesins in the *in vitro* gliding assay). In the experiments, the parameters that were considered in loading and transporting of oligonucleotides-linked microbeads will be varied in order to successfully unload oligonucleotides-linked microbeads. The considered parameters include the density of oligo-nucleotides, the length of the base sequence of oligo-nucleotides, and the gliding speed of oligonucleotides-linked microtubules.

The loading/transporting/unloading processes may be observed through a fluorescence microscope by labeling the gliding microtubules and microbeads with different colors of fluorescent dyes.

## 4. CONCLUSIONS

This paper presents a design of an autonomous molecular propagation system that loads, transports, and unloads information molecules using DNA hybridization and biomolecular linear motors. The authors of this paper are currently conducting biochemical experiments to empirically examine the system design.

The autonomous molecular propagation system described in this paper is applicable to communication among biological nanomachines using molecules as communication carrier. It would also enable a number of new applications in the bio-nanotechnology. For instance, the autonomous molecular propagation system may extend and enhance the existing molecular transporting mechanisms (e.g., microfluidics) in micro total analysis systems (micro TAS) or lab-on-a-chip [14] by providing new means to deliver single molecular samples and reagent between components (such as biochemical sensors and reactors) on a chip. With such an advanced micro TAS, it will become possible to conduct biochemical analysis and synthesis (such as cell analysis and blood diagnosis) on a small chip in a new manner.

## 5. ACKNOWLEDGEMENTS

The authors of this paper would like to thank Associate Professor Yoko Y. Toyoshima and Mr. Kenya Fruta of the Department of Life Sciences, the University of Tokyo for helping with biological preparation.